\documentclass[twocolumn,secnumarabic,amssymb, nobibnotes, aps, prl]{revtex4-1}
\usepackage{amsmath,graphicx,float}

\begin{document}
\bibliographystyle{unsrt}

\title{Brownian motion: from kinetics to hydrodynamics}%

\author{Hanqing Zhao$^{1,2}$,Hong Zhao$^{1,3}$}
\email{zhaoh@xmu.edu.cn}
\address{$^1$Department of Physics and Institute of Theoretical Physics and Astrophysics, Xiamen University, Xiamen 361005, China}
\address{$^2$Department of Modern Physics, University of Science and Technology of China, Hefei 230026, China}
\address{$^3$Collaborative Innovation Center of Chemistry for Energy Materials, Xiamen University, Xiamen 361005, China}

\date{january 18, 2017}%
\begin{abstract}
 Brownian motion has served as a pilot of studies in diffusion and other transport phenomena for over a century. The foundation of Brownian motion, laid by Einstein, has generally been accepted to be far from being complete since the late 1960s, because it fails to take important hydrodynamic effects into account. The hydrodynamic effects yield a time dependence of the diffusion coefficient, and this extends the ordinary hydrodynamics. However, the time profile of the diffusion coefficient across the kinetic and hydrodynamic regions is still absent, which prohibits a complete description of Brownian motion in the entire course of time. Here we close this gap. We manage to separate the diffusion process into two parts: a kinetic process governed by the kinetics based on molecular chaos approximation and a hydrodynamics process described by linear hydrodynamics. We find the analytical solution of vortex backflow of hydrodynamic modes triggered by a tagged particle. Coupling it to the kinetic process we obtain explicit expressions of the velocity autocorrelation function and the time profile of diffusion coefficient. This leads to an accurate account of both kinetic and hydrodynamic effects. Our theory is applicable for fluid and Brownian particles, even of irregular-shaped objects, in very general environments ranging from dilute gases to dense liquids. The analytical results are in excellent agreement with numerical experiments.
\end{abstract}
\maketitle

 The subject of particle diffusion in a fluid, with Brownian motion as a
 canonical example, play vital roles in various fields.
 The state-of-the-art technologies have made accurate measurements of instantaneous position and even velocity of individual molecules within the experimental
 reach.
 Owing to this, in recent years there have been a renewal of experimental
 interests in reexamining the diffusion theory \cite{1,2,3,4,5,6,7,8,9,10,11,12,13,14,15,16,17,18,19,20,21,a1}, with a particular attention paid to diffusion anomalies in a number of systems, ranging from passive tracers\cite{7,8,9,10,11,12,13,14,15,16,17} to active matters\cite{18,19,20,21}. These anomalies are generally conceived to be of hydrodynamic origin. They do not fall into the canonical paradigm\cite{22,23} of Brownian motion, which assumes an exponential decay of velocity autocorrelation function (VACF). Contrary to this assumption, hydrodynamics is well known to result in the recovery of memory of particle velocity and eventually lead to a very slow decay of VACF\cite{24,25,26,27,28,29,30,31,32,33,34,35,36,37}. Theoretical progrespses achieved so far\cite{22,23,24,25,26,27,28,29,30,31,32,33,34,35,36,37,38,39,40,41,42,43}, though able to capture hydrodynamic effects to some extent, are restricted to some limiting cases, and results obtained from different approaches are controversial. The hydrodynamic effects yield a time dependence of the diffusion coefficient, contrary to the assumption that the diffusion coefficient is a constant in the entire time course. A complete solution thus should be worked out in the framework of generalized hydrodynamics\cite{42,43}, which crucially differs from the ordinary one in that the transport coefficients could have a time dependence. Such a framework provides a way for going beyond the continuum approximation, as done in ordinary hydrodynamics, and entering into the regime that the discrete feature of particles plays a role. What is the time profile of the diffusion coefficient? This is an important question which remains open. In addition, previous hydrodynamic approaches aim at the diffusion of a fluid particle. To unify the diffusion of fluid particles and Brownian particles with arbitrary shapes is important to establish a complete theory of diffusion. These are the main issues to be addressed in this paper.

Before exposing the details of our approaches, it may be useful to sketch the technical idea here. As usual, we divide the VACF into the kinetic and hydrodynamic parts, respectively. However, the separation introduced here is dramatically different from what has been done previously\cite{26,27,31}. Most importantly, the kinetic part refers to the simplest kinetic processes corresponding to the molecular chaos approximation. It accounts only for the process that momentum initially carried by the tagged particle is transferred out of the tagged particle.  In essence, the kinetic (hydrodynamic) part is responsible for the memory loss (recovery) of velocity, and they have no overlap as appearing in previous theories.
 As such, this definition of kinetics excludes the ring collision. The transferred momentum triggers a vortex backflow in fluid. Then, we solve linearized hydrodynamic equations to obtain the shear viscosity and sound modes\cite{29}, which in turn give the analytical solution of the backflow\cite{25}. Noticing that the memory recovery solution, for the backflow eventually drives the tagged particle to move with it, we obtain two coupled equations satisfied by VACF and the diffusion coefficient. The solution to these equations immediately yield the formula of the VACF and the time profile of the diffusion coefficient.

We will derive a set of equations satisfied by the VACF and the diffusion coefficient. The former is defined as $C(t)=\frac{1}{d}\langle\mathbf{p}(t)\cdot\mathbf{p}(0)\rangle/M^{2}$, where $\mathbf{p}(t)$ is the momentum of a tagged particle at time t, $M$ is its mass, $d$ is the dimension of the fluid, and $\langle\cdot\rangle$ represents the ensemble average. The tagged particle can be either a fluid or a Brownian particle. The diffusion coefficient is related to the VACF via the Green-Kubo formula. The instant motion of tagged particle can be decomposed into three, dubbed the kinetic, hydrodynamic and random parts, respectively. As the tagged particle collides with fluid particles, it transfers part of the momentum to the latter. The kinetic part, denoted as $\mathbf{p}^{K}(t)$, is not transferred.  The hydrodynamic part, denoted as $\mathbf{p}^{H}(t)$, refers to the momentum transferred to the surrounding fluid particles. It relaxes according to hydrodynamics, and is transferred back to the tagged particle. The random part, denoted as $\mathbf{p}^{R}(t)$, is uncorrelated with the initial momentum and comes from random collisions with other particles. Taking these into account, we have
\begin{equation}
\mathbf{p}(t)=\mathbf{p}^{K}(t)+\mathbf{p}^{H}(t)+\mathbf{p}^{R}(t).\label{1}
\end{equation}
 Correspondingly, we have $C(t)=C_{K}(t)+C_{H}(t)$. Note that $\mathbf{p}^{R}(t)$ has no correlation with the initial momentum. From the kinetic theory we have $C_{K}(t)=C(0)\exp(-\frac{C(0)}{D_{0}}t)$\cite{23}, and thus $\mathbf{p}^{K}(t)=\mathbf{p}(0)\exp(-\frac{C(0)}{D_{0}}t)$, where $D_0$ is the kinetic diffusion constant of tagged particle and $C(0)=\frac{k_{B}T}{M}$($k_{B}$ is the Boltzmann constant and $T$ the temperature).

 Next, we calculate $\mathbf{p}^{H}(t)$. Let the probability for the tagged particle to appear in a unit volume at $\mathbf{r}$  and $t$ be $\rho(\mathbf{r},t)$, and the density of momentum transferred into this volume be $\mathbf{p}(\mathbf{r},t)$. Note that each particle in this volume has an average velocity $\mathbf{p}(\mathbf{r},t)/\rho$, namely, the velocity field of vortex backflow triggered by the tagged particle. Here $\rho$ is the density of fluid. The tagged particle, irrespective of its shape and mass, has the same velocity as the vortex backflow on average since it is driven by the backflow. As a result,
\begin{equation}
\mathbf{p}^{H}(t)=\frac{M}{\rho}\int\mathbf{p}(\mathbf{r},t)\rho(\mathbf{r},t)d\mathbf{r}\label{2}.
\end{equation}

 To calculate $\mathbf{p}(\mathbf{r},t)$, we employ the hydrodynamic approach\cite{31}. Without loss of generality we assume that initially the tagged particle
 is at the origin with a momentum $\mathbf{p}(0)\equiv[p_{x}(0),0,0]$. Then, $\mathbf{p}(\mathbf{r},t)\equiv[p_{x}(\mathbf{r},t),0,0]$ because of the momentum conservation. $p_{x}(\mathbf{r},t)$ follows hydrodynamic equations which arise from the conservation laws of the particle number, energy and momentum. We then linearize these equations and perform the Fourier (Laplace) transformation with respect to space (time). This procedure is rather standard. The crucial difference of our treatments from others is that we set the initial conditions to $\mathbf{p}(\mathbf{r},t=0)=[p_{x}(0)\delta(\mathbf{r}),0,0]$, $\Delta T(\mathbf{r,}t=0)=\Delta T\delta(\mathbf{r})$, and $\Delta n(\mathbf{r,}t=0)=\Delta n\delta(\mathbf{r})$, where $\Delta T(\mathbf{r,t})$ and $\Delta n(\mathbf{r,t})$ are the temperature and particle density fluctuations, respectively. It is precisely due to this initial condition that the $5\times5$ hydrodynamic matrix is reduced into a $3\times3$ one, and subsequent analytical derivations are substantially simplified. Solving the hydrodynamic equations (see Supplementary Materials\cite{a2}, section S1), we obtain $p_{x}(\mathbf{k},t)=p_{x}^{V}(\mathbf{k},t)+p_{x}^{S}(\mathbf{k},t)$ under the long wave approximation, where
 \begin{equation}
\frac{p_{x}^{V}(\mathbf{k},t)}{p_{x}(0)}=\frac{k_{y}^{2}+k_{z}^{2}}{k^{2}}\exp(-\nu_{0}k^{2}t)\label{3}
\end{equation}
is the contribution of the shear viscosity mode and
\begin{equation}
\frac{p_{x}^{S}(\mathbf{k},t)}{p_{x}(0)}=\frac{k_{x}^{2}}{k^{2}}\exp(-\Gamma k^{2}t)\cos(c_{s}kt)\label{4}
\end{equation}
of the sound mode. Here $\nu_{0}$ is the kinetic viscosity diffusivity, $c_{s}$ the sound speed and $\Gamma$ the sound attenuation coefficient. Note that these solutions apply to a two-dimensional fluid as well, for which $k_{z}=0$. These solutions are anisotopic, implying that the profile of backflow triggered by the tagged particle is anisotopic.

From the kinetic theory we find $\rho(\mathbf{r},t)=\frac{1}{4\pi D_{0}t}\exp(-\frac{r^{2}}{4D_{0}t})$. When the memory effect is taken into account, the diffusion coefficient is in general time dependent. This is a key ingredient of the generalized hydrodynamics. We hypothesize that $\rho(\mathbf{r},t)$ has the same form, except that $D_{0}$ is replaced by $D_{H}(t)+D_{0}$ (see Supplementary Materials\cite{a2}, section S2). This gives $\rho(\mathbf{k},t)=\exp[-(D_{H}(t)+D_{0})k^{2}t]$. [In principle, the kinetic viscosity diffusivity should also be replaced by $\nu_H(t)+\nu_0$  following the generalized hydrodynamics. Nevertheless, our molecular dynamic simulations, as well as previous studies\cite{38} show that this coefficient has a negligible time dependence. Thus, this parameter is set to be time-independent throughout.]

To solve Eq.(2) we first employ the Parseval formula to transform the spatial integral into a wave-vector integral, i.e., $\int p_{x}(\mathbf{r},t)\rho(\mathbf{r},t)d\mathbf{r}=(\frac{1}{2\pi})^{d}\int p_{x}(\mathbf{k},t)\rho(\mathbf{k},t)d\mathbf{k}$. Notice that at time t, the momentum transferred to the surrounding fluid particles is $p_{x}(0)[1-\exp(-\frac{C(0)}{D_{0}}t)]$, while the remainder, i.e., $p_{x}(0)\exp(-\frac{C(0)}{D_{0}}t)$, is still carried by the tagged particle. To determine the time after which the backflow triggered by this part of momentum sets in and thus Eq. (2) applies, we introduce a function $R(t)$ to characterize retarded effects in response. Taking these into account, for the correlation $C_H(t)=\langle p_x^H(t)p_x(0) \rangle$, we have
\begin{eqnarray}
\frac{C_H(t)}{C(0)}& = & \frac{M(d-1)}{\rho d}(1-e^{-\frac{C(0)}{D_{0}}t})R(t)\nonumber \\
 &  & [4\pi(D_{H}(t)+D_{0}+\nu_{0})t]^{-\frac{d}{2}},
\end{eqnarray}
where we neglect the contribution of sound mode. [The sound mode may contribute a small correction to VACF, however, negligible usually (see Supplementary Materials\cite{a2}, section S3).] On the other hand, by the definition $D(t)=\frac{1}{2d}\frac{d}{dt}\langle r^{2}(t)\rangle$, one has $D(t)=\int_{0}^{t}C(t^{\prime})dt^{\prime}$, which gives the usual Green-Kubo formula at $t\rightarrow\infty$. Inserting $C_K(t)=C(0)\exp(-C(0)t/D_0)$ into the formula, we have $D(t)=D_0(1-\exp(-C(0)t/D_0))+D_H(t)$, with
\begin{equation}
D_H(t)=\int_{0}^{t}C_H(t^{\prime})dt^{\prime}.
\end{equation}
Solving the coupled Eqs. (5) and (6), we obtain
 \begin{eqnarray}
\frac{C_{H}(t)}{C(0)} & = & \frac{M(d-1)}{\rho d}(1-e^{-\frac{C(0)}{D_{0}}t})R(t)\nonumber \\
 &  & (4\pi t)^{-\frac{d}{2}}A^{-\frac{d}{d+2}},\\
D_{H}(t) & = & A^{\frac{2}{d+2}}-(D_{0}+\nu_{0})
\end{eqnarray}
 with
\begin{eqnarray}
 A&=&(D_{0}+\nu_{0})^{\frac{d+2}{2}}+\frac{M(d-1)(d+2)}{2\rho d}(4\pi)^{-\frac{d}{2}}C(0)\nonumber\\
 & &\int_{0}^{t}(1-e^{-\frac{C(0)}{D_{0}}t^{\prime}})R(t^\prime)t^{\prime-\frac{d}{2}}dt^{\prime}
\end{eqnarray}

Now we find the explicit form of $R(t)$. First of all, provided we assume that the momentum transferred out of the tagged particle initializes a backflow immediately (that is, there is no retarded effects), we have $R(t)=1$. Our numerical simulations show that this result works perfectly at low densities, but has inaccuracies at higher densities. In  addition, such a response function may cause $C_H(t)$ to diverge at $t\rightarrow0$, which is unphysical and inconsistent with the kinetic theory.  To find the form of $R(t)$ we note that the recovery of momentum memory takes place after a ring collision\cite{26,27}. The lowest order rings involve three collision events. This process defines the time threshold at which the backflow begins to play a role. The free time obeys the Gamma distribution, i.e., $\frac{1}{\tau} \exp(-t/\tau)$, where $\tau$ is the mean free time. Since $R(t)$ is the probability for the memory recovery to occur, it is the joint probability for three independent collision events to occur. This gives
\begin{eqnarray}
R(t)&=&1-\frac{\tau_B^2}{(\tau_F-\tau_B)^2}\exp(-t/\tau_F)\nonumber \\
    & & -((\tau_B-\tau_F)t+2\tau_F\tau_B-\tau_F^2)\exp(-t/\tau_F),
\end{eqnarray}
where $\tau_B$ is the mean inverse frequency of the Brownian particle-fluid particle collisions, and $\tau_F$ of collisions between fluid particles. When the tagged particle is a fluid particle, Eq. (10) reduces to
\begin{equation}
R(t)=1-\frac{2\tau_F^2+2\tau_Ft+t^2}{2\tau_F^2}\exp(-t/\tau_F).
\end{equation}
In principle, $R(t)$ should take higher order ring collisions as well as other factors into acount, but these factors are unimportant as long as the leading hydrodynamic contribution is concerned.
\begin{figure}[t]

\includegraphics[width=8.7cm,height=12.4cm]{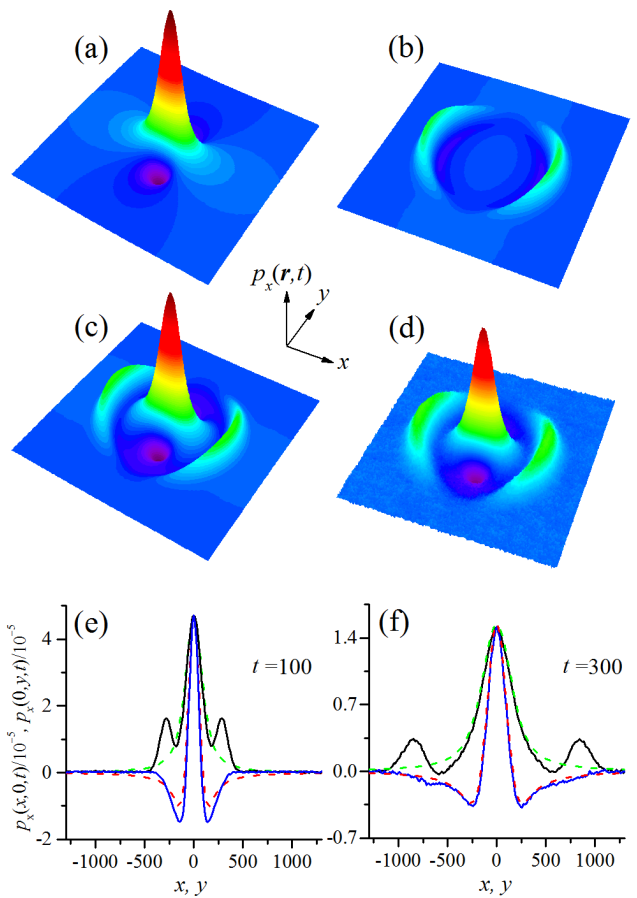}

\caption{The hydrodynamic modes. The analytical results for the viscosity mode (a), the sound mode (b), and their superposition (c) at $t=300$ for $\nu_{0}=8.0$. (d) The simulated $p_{x}(\mathbf{r},t)$ at $t=300$; (e) and (f) The intersections of the simulated $p_{x}(\mathbf{r},t)$ with $y=0$ (black solid lines) and $x=0$ (blue solid lines) at $t=100$ and $300$, respectively. They are compared with the viscosity mode given by Eq. (12) with the best fitting value of $\nu_{0}=8.0$.
 The intersections of the profiles thereby obtained with $y=0$ ($x=0$) are represented by the green dashed (red dashed) lines. For simulations $\sigma=6$.}

\end{figure}
 We use the hard-disk model to simulate fluids. It consists of $N$ disks of unit mass moving in a square box of size $L$. Moreover, the periodic boundary condition is used. For simulation results to be free from finite-size effects at $t<L/(2c_{s})$\cite{39}, we set $L=2000$(see Supplementary Materials\cite{a2}, section S4). The particle number $N=40000$, corresponding to an average disk number density $n=0.01$. The packing density $\phi=n\pi\sigma^{2}/4$, with $\sigma$ being the disk diameter. The system's behavior at $2\leq\sigma\leq9$, which covers from the gas to the liquid regime, are studied in great details. Note that for this model the crystallization density is $\phi=0.71$, corresponding to $\sigma=9.5$. We simulate the system by the event-driven algorithm. The rescaled temperature $T=1$($k_{B}$ is set to be unity.).

 To proceed, we note that the viscosity mode, the inverse Fourier transform of Eq. (3) is
\begin{equation}
\frac{p_{x}^{V}(\mathbf{r},t)}{p_{x}(0)}=\frac{x^{2}-y^{2}}{2\pi r^{4}}(1-e^{-\frac{r^{2}}{4\nu_{0}t}})+\frac{y^{2}}{4\pi r^{2}\nu_{0}t}e^{-\frac{r^{2}}{4\nu_{0}t}}.\label{11}
\end{equation}
The sound mode can be obtained by numerically performing the inverse transform of Eq. (4). Combining them together leads to a theoretical prediction for $p_{x}(\mathbf{r},t)$. As shown in Fig. 1(a)-(c), $p_{x}(\mathbf{r},t)$ is anisotopic. Note that Eq. (12) gives the $x$-component of velocity field of vortex backflow. The $y$-component of velocity field can also be obtained similarly. Therefore, the viscosity mode gives the analytical solution of the conventional vortex backflow.

\begin{figure}[t]

\centerline{\includegraphics[width=8.7cm,height=8.7cm]{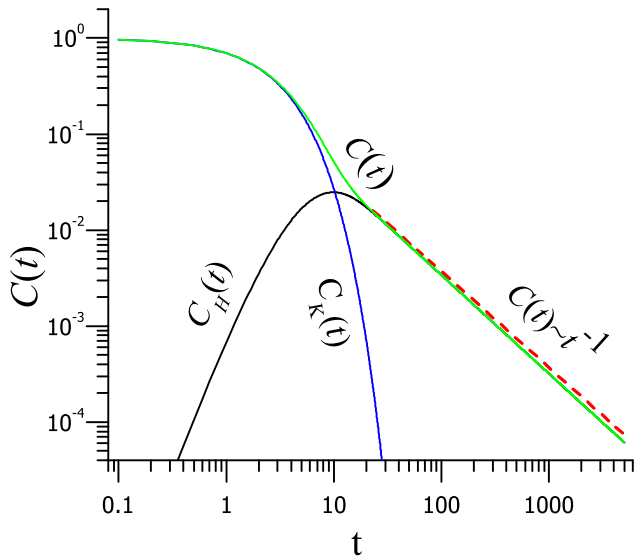}}
\caption{The contributions of $C_K(t)$ and $C_H(t)$, and their combination C(t). $C_K(t)$ is calculated by $C_K(t)=C(0)\exp(-C(0)t/D_0)$ and $C_H(t)$ is obtained by  Eq. (7). For simulations $\sigma=6$. }

\end{figure}

Numerically, $p_{x}(\mathbf{r},t)$ is computed by the correlation function $\langle\widetilde{p}_{x}(\mathbf{r},t)p_{x}(0)\rangle$, i.e.,
\begin{equation}
\frac{p_{x}(\mathbf{r},t)}{p_{x}(0)}=\frac{\langle\widetilde{p}_{x}(\mathbf{r},t)p_{x}(0)\rangle-\langle\widetilde{p}_{x}(\mathbf{r}\neq0,0)p_{x}(0)\rangle}{\langle|p_{x}(0)|^{2}\rangle}.\label{12}
\end{equation}
Here $\widetilde{p}_{x}(\mathbf{r},t)$ represents the $x$-component of the instantaneous momentum density. The result is shown in Fig. 1(d), which confirms the theoretical prediction. The explicit expression of the viscosity mode allows us to compute $\nu_{0}$ based on the simulated $p_{x}(\mathbf{r},t)$. Specifically, we fit the viscosity mode, i.e., the center peak, of the simulated $p_{x}(\mathbf{r},t)$ by $p_{x}^{V}(\mathbf{r},t)$. In this way, we find $\nu_{0}=8.0$ for $\sigma=6$. Note that this value is independent of time [c.f. Fig. 1(e)-(f)], implying that hydrodynamics is harmless to the viscosity diffusivity. Previous numerical studies based on the Helfand-Einstein formula have shown that this diffusivity is independent of the system size either\cite{38}, consistent with present findings. In Table 1 we summarize the value of $\nu_{0}$ computed for various packing densities. It is important to note that the result is close to that given by the Enskog formula under the first Sonine polynomial approximation\cite{36} in the dilute gas regime, but the discrepancy between these two results becomes significant as the packing density increases (see Table 1). [The state equation used in the Enskog formula is the Henderson expression\cite{38}].

 The value of $D_0$ can be determined by fitting the simulated VACF using our analytical result. To be specific, in the short time regime hydrodynamics does not effect and $C(t)=C(0)\exp(-C(0)t/D_0)$. Using this formula to fit the VACF we obtain $D_0$ numerically (see Supplementary Materials\cite{a2}, section S5). The results for fluid particles are given in Table 1. Alternatively, $D_0$ can be found by the Enskog formula, and the results are given in Table 1 also. The results for $D_0$ obtained by these two different methods are in excellent agreement for almost all densities: the discrepancy is less than $10\%$ even for the highest density ($\sigma=9$) available in simulations. Usually, it is believed that the Enskog formula is  accurate at low densities\cite{31}. Our result shows that, contrary to this common wisdom, it is valid at much broader regime, if we interpret it as the full kinetic part of diffusion coefficient. There is one more way to calculate $D_0$. That is, one can use the relation between the mean free time $\tau$ and the kinetic diffusion constant\cite{31}, i.e., $\tau=\frac{d}{2}\frac{D_0}{C(0)}$. The parameter $\tau$ can be easily measured numerically, and the result is given in Table 1. The results are in perfect agreement with those obtained by fitting the short time behavior of VACF.

With $D_0$, $\tau$ and $\nu_0$ at hand, we can calculate the kinetic and hydrodynamic parts of VACF, i.e., $C_K(t)$ and $C_H(t)$. Figure 2 shows how they affect the overall behavior of VACF at a moderated packing density for the hard-disk model. In this case, $D_0$ is relatively large. As a result, $C_K(t)$ decays slowly, and appears in a relatively  large time interval. Although the peak of $C_H(t)$ shows up in this interval, the amplitude is small. So, in the total VACF, $C_K(t)$ dominates over $C_H(t)$ in this interval. Thus, the VACF exhibits  an exponential decay followed by a power-law tail. The above analysis suggests that depending on the quantitative profiles of $C_K(t)$ and $C_H(t)$, the total VACF can exhibit rich behavior in the crossover region, when the kinetic parameters vary. In particular, for sufficiently small $D_0$, for $t$ much smaller than the time corresponding to the peak of $C_H$, $C_K(t)$ may be much smaller than the peak  magnitude of $C_H$. In this case, the VACF could exhibit a peak between the stages of exponential and power law decays.  [In principle, one should take the sound mode into account. However, its contributions are negligibly small (see Supplementary Materials\cite{a2}, section S3).]

\begin{figure*}[t]
\begin{center}
\centerline{\includegraphics[width=17.8cm,height=7.82cm]{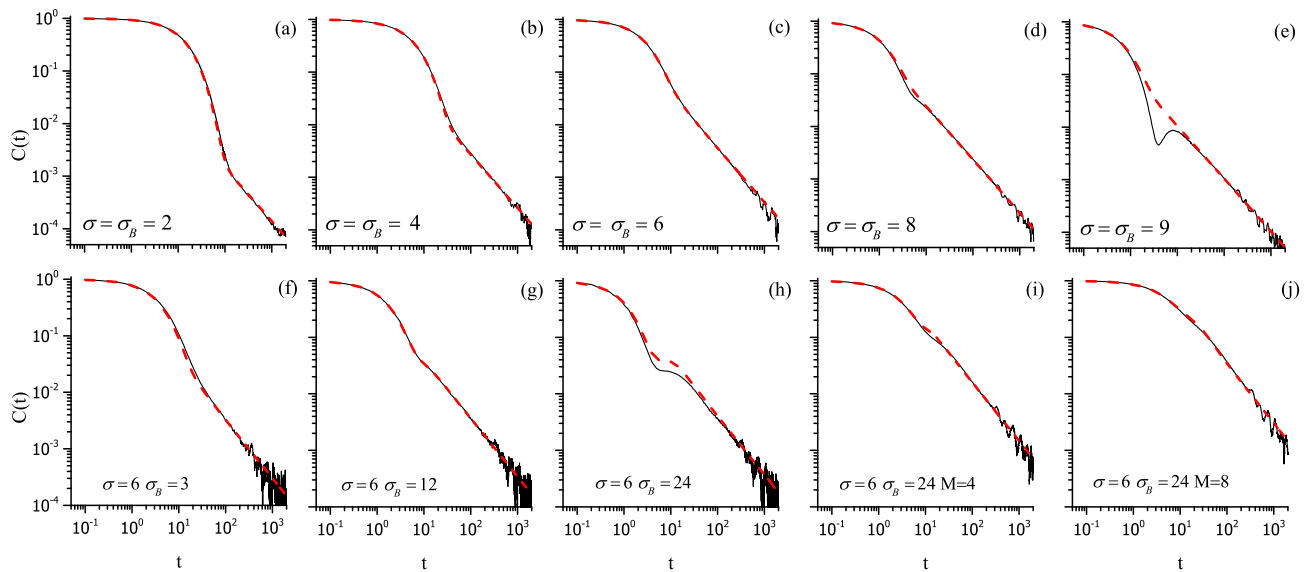}}

\caption{Comparison between theoretical and simulation results of VACF. (a)-(e) Simulated (black solid lines)
 and analytical (red dashed lines) VACF at various densities.
 (f)-(h) for a Brownian particle with diameter $\sigma_B$=3, 12 and  24, respectively, in a fluid with particle diameter $\sigma=6$. For (a)-(h), the mass of the tagged particle is set to unity as fluid particles. (i)-(j) for a Brownian particle with diameter $\sigma_B$= 24 and mass M=4 and 8, respectively, in fluid with $\sigma=6$. The parameters $\nu_0$ and $D_0$ for fluid particles are given in Table 1. $D_0$ for Brownian particles in (f)-(j) is obtained by fitting the short-time behavior of simulated VACF at $t\rightarrow 0$ in the way introduced (see SI Appendix section S5),this gives $D_0=3.85,1.70,1.1,0.85,0.80$, corresponding to (f)-(j).
     }
\end{center}
\end{figure*}

Figure 3(a)-(e) show the VACF for a fluid particle at different packing density, Fig. 3(f)-(h) are for a Brownian particle of unit mass with different diameters, and Fig. 3(i)-(j) are for a Brownian particle of fixed diameter but with different masses. We see that in most cases Eq. (7) is in excellent agreement with the simulation results. In particular, for Brownian particles with large mass, as usual situations where Brownian particle is much bigger than fluid particles and thus have a larger mass, the agreement is perfect. In certain extreme conditions, e.g., the fluid density close to the crystallization density or Brownian particle with small mass but big size, there is a small discrepancy between theoretical and numerical results in the crossover region, as shown in Fig. 3(e) and (h). These situations are nevertheless extremely rare. Indeed, for Fig. 3(h) the diameter of the Brownian particle is four times larger than a fluid particle, but its mass is still the same as that of a fluid particle. Even in such extreme cases, the deviation in the diffusion coefficient is negligible in a relative large time regime in two-dimensional fluid (see Supplementary Materials\cite{a2}, section 6). It is possible that the deviation can be fixed by a more accurate response function $R(t)$.

We first compare our results to previous theories. Let us compare our result with those obtained by the kinetic theory, the traditional hydrodynamic approach, the self-consistent mode coupling approach\cite{40,41}, and the generalized Langevin equation approach\cite{32,33,34,35,36,37}. We first consider the short time regime. The kinetic theory applies in this regime. The generalized Langevin equation approach can give a closed form of VACF, and thus, presumably, applies in this regime also.  We have shown that $C_K(t)$ obtained by the kinetic theory can accurately describe the VACF at sufficiently short time even at high densities (see Supplementary Materials\cite{a2}, section S5). From Eqs. (7) and (8), as well as Fig.2, we see that at short time kinetic contributions dominate over hydrodynamic ones. In other words, $C_{H}(t)$ and $D_{H}(t)$ are negligible and diffusive dynamics is described by the kinetic theory. Therefore, our theory is consistent with the standard kinetic theory. In contrast, the generalized Langevin equation fails to achieve this. This method, besides the empirical nature of the  kernel function introduced to characterize the memory effect, assumes incompressiblity that leads to an unphysical consequence, i.e., $C(t\rightarrow0)=\frac{k_{B}T}{M_{\ast}}$, contradictory to the result of $C(t\rightarrow0)=\frac{k_{B}T}{M}$ given by the kinetic theory\cite{23} and to the equipartition theorem of statistical physics\cite[35,36], where $M_{\ast}=M+\frac{2}{3}\pi R^{3}\rho$ is the effective mass arising from the assumption of incompressiblity.

\begin{table*}[htbp]
\begin{tabular}{cccccc}
\hline
$\sigma(\phi)$&2(0.03)&4(0.13)&6(0.28)&8(0.50)&9(0.63)\\
\hline
$\nu_{0}$(E)&14.1&7.7&6.3&8.5&14.3\\
$\nu_{0}$(S)&$14.3\pm0.05$&$8.3\pm0.05$&$8.0\pm0.05$&$15.0\pm1.0$&$35.0\pm2.0$\\
$D_{0}$(E)&13.40&5.70&2.76&1.10&0.59\\
$D_{0}$(S)&$13.35\pm0.02$&$5.68\pm0.02$&$2.74\pm0.02$&$1.14\pm0.02$&$0.66\pm0.02$\\
$\tau(S)$&$13.24\pm0.03$&$5.64\pm0.0$3&$2.74\pm0.03$&$1.14\pm0.03$&$0.65\pm0.03$\\
\hline
\end{tabular}
\caption{Kinetic coefficients obtained by the Enskog formula (E) and by simulations (S).}
\end{table*}

 At the long time limit, the kinetic theory is not applicable. There exist theories describing the long time behavior. Now we compare our results with the predictions of these theories. For d=2, Eqs. (7)-(8) give the asymptotic solution, $C_{H}(t)=C(0)\sqrt{\frac{M}{16\pi\rho}}(t\sqrt{\ln t})^{-1}$ and $D_{H}(t)=C(0)\sqrt{\frac{M\ln t}{4\pi\rho}}$. They are in agreement with the results obtained by the self-consistent mode coupling approach, but contradictory to the $t^{-1}$ law predicted by the traditional hydrodynamic approach. To understand this discrepancy we see from Eq. (5) that the asymptotic solutions apply for $t\gg t_{c}\equiv\exp[4\pi\rho M(\frac{D_{0}+\nu_{0}}{k_{B}T})^{2}]$, when ($D_{0}+\nu_{0}$) is much smaller than $D_{H}(t)$. For sufficiently large $t$ but smaller than $t_{c}$, when hydrodynamic contributions have dominated already (i.e.,$(1-e^{-\frac{C(0)}{D_{0}}t})R(t)\sim1$) but $D_{H}(t)$ is still negligible, the $\sim t^{-1}$ behavior shows up. That is, the $\sim t^{-1}$ behavior appears in intermediate time, and crosses over to $\sim(t\sqrt{\ln t})^{-1}$ behavior at $t_{c}$. Therefore, our theory unifies the traditional hydrodynamic and self-consistent mode coupling approaches. With the kinetic coefficients at hand we can estimate the transient time $t_{c}$, which is $10^{23}$, $10^{10}$, $10^{4}$, $10^{13}$, and $10^{35}$, respectively, for the packing densities listed in Table 1. The value of $t_{c}$ is large at the dilute gas and dense fluid limits because in the former (latter) $D_{0}$ ($\nu_{0}$) is large. Since the time interval ($t\sim10^{3}$) considered here is much shorter than $t_{c}$, the asymptotic behavior $C(t)\sim t^{-1}$ shows up in both cases. We thus expect that in usual simulations or experimental measurement the tail of $C_H(t)$ follows a $t^{-1}$ law. The deviation from the $t^{-1}$ law may be feasible (cf. Fig. 2), but is not significant in general. In this sense, though the traditional hydrodynamic approach is not exact at the thermodynamic limit, it is a good approximation practically.

 For $d=3$, our theory gives $C(t)/C(0)\sim\frac{2}{3}\frac{M}{\rho}[4\pi(D_{0}+\nu_{0}+D_{H}(\infty))t]^{-\frac{3}{2}}$ at the long time limit. Here $D_{H}(\infty)$ is the saturation value of $D_{H}(t)$. $D_{H}(\infty)=[(D_{0}+\nu_{0})^{\frac{5}{2}}+0.12\frac{k_BT}{\pi \rho}(\frac{C(0)}{D_0})^\frac{1}{2}]^{\frac{2}{5}}-(D_{0}+\nu_{0})$
  for a fluid particle ($M=m$). (For a Brownian particle, the expression can be obtained also, though has a more complicated form.) This asymptotic behavior differs also from previous ones. Indeed, for the fluid particle, previous hydrodynamic approaches\cite{24,25,26} give $\frac{C(t)}{C(0)}\sim\frac{2}{3}\frac{M}{\rho}[4\pi(D_{0}+\nu_{0})t]^{-\frac{3}{2}}$, while for the Brownian particle, the generalized Langevin equation approach gives $\frac{C(t)}{C(0)}\sim\frac{2}{3}\frac{M_{\ast}}{\rho}(4\pi\nu_{0}t)^{-\frac{3}{2}}$\cite{33,37}. The former corresponds to the neglecting of $D_{H}(\infty)$ in our results. The latter deviates significantly from ours where there is no room for $D_{0}$, and $m$ replaced by the effective one $M_\ast$ leading to the enhancement of C(t). We see that, the result obtained by the generalized Langevin equation is a crude approximation of ours(see Supplementary Materials\cite{a2}, section 7).

 Next we discuss the hydrodynamic effects at macroscopic scales. Generally speaking, the power-law tail leads to a divergent diffusion coefficient in the thermodynamic limit in two-dimensional fluids, according to Eq. (8). Practically, it is important how it influences the behavior of a real system of finite size. To address this issue we note that the average distance between two neighboring molecules in the air is about $10^{-9}$ meters. Corresponding to one centimeter, we have $L\sim10^{8}$ for our model and therefore the time within which a particle diffuses freely without being influenced by the boundaries is $\sim10^{7}$. Using Eq. (8), we estimate $D_H(t = 10^7)/D_0 \sim 0.1, 0.7, 2.1, 7.3, 7.3$ for fluid particles in the hard-disk model with $\sigma=2,4,6,8,9$, respectively. We see that in a dilute gas the kinetic contribution dominates, but the hydrodynamic contribution increases with the packing density and eventually dominates over the kinetic contribution.

 In a three-dimensional fluid the hydrodynamic effect is much weaker usually. Using the kinetic coefficients of  hard-sphere model, we estimate $D_{H}(\infty)/D_{0}<1\%$ for $\sigma\leq9$ for a fluid particle. For a Brownian particle, $D_0$ may decrease with the increase of diameter (See Fig. 3). Supposing a Brownian particle has a kinetic diffusion constant of $D_0\sim0.1$, we estimate that the hydrodynamic contribution can reach $D_H(\infty)/D_0\sim10\%$. With the further decrease of $D_0$, the hydrodynamic contribution to the diffusion coefficient may become to dominate over the kinetic contribution. As a manifestation, the tagged particle is driven by the backflow, moving forward instead of exhibiting kinetic diffusion. However, since $C_{H}(t)\sim t^{-3/2}$ at large time, the convergence to the saturation value $D_H(\infty)$ is very fast, occurring at a microscopic time scale of $\frac{C(0)}{D_{0}}$. Due to this one must be cautious in identifying the hydrodynamic effect in a three-dimensional fluid. The time scale of experimental measurement of the diffusion coefficient is usually macroscopically large, much larger than the time scale for the hydrodynamic contribution to saturate. In such a case, the measured diffusion coefficient includes a hydrodynamic contribution, which might be even bigger than the kinetic contribution. Where this hydrodynamic origin not unveiled, one may incorrectly attribute the measured value to the normal kinetic diffusion constant.

Last, we argue that generalized linear hydrodynamics is sufficient for capturing hydrodynamic effects. We see that in Eqs. (7) and (8) only kinetic constants are input parameters. Once they are given, the hydrodynamic contributions are fully determined. The excellent agreement between the theory and the numerical confirmation implies that the generalized linear hydrodynamics suffices to take the hydrodynamic effect into full account, at least for
the hard-disk fluid. This is also true for Brownian particles of arbitrary shape. The reason is that the hydrodynamic effect results from the backflow of fluid. Once the kinetic parameter $D_0$ of the Brownian particle is obtained, according to Eqs. (7) and (8), both the
VACF and the diffusion coefficient are completely determined. Therefore, to find hydrodynamic contributions accurately relies on the accurate calculation of kinetic parameters such as kinetic diffusion constant, the
mean free time, the kinetic viscosity diffusivity, and the response function.

As expected by the generalized hydrodynamics, in general, the viscosity diffusivity is also time independent, and it should be coupled to Eq. (5). However, as mentioned before, based on our results as well as previous studies it is at most weakly time-dependent. The hydrodynamic correction to the diffusion coefficient resulting from the time profile (if any) of the viscosity diffusivity should be of higher order.

We present the first quantitative results suitable for describing the crossover of Brownian motion from kinetic to hydrodynamic process. Our theory consists of a set of self-consistent equations satisfied by the VACF and the diffusion coefficient. This allows us, for the first time, to describe accurately the kinetic-hydrodynamic crossover behavior of VACF and diffusion coefficient. As hydrodynamic memory effects are due to the vortex backflow that drives the Brownian particle, our theory applies a general Brownian particle, regardless of its shape and size. The result of the generalized Langevin equation is a crude approximation of our formula in the limit of $D_0\rightarrow0$.     As such, our theory is not only important for experimental studies, but also applies to a broad spectrum of subjects.

Our theory paves a way for solving some long-standing transport problems, notably anomalous heat transport for which hydrodynamics is commonly conceived to be responsible. Interestingly, in spite of the classical nature of phenomena studied here, our theory bears a firm analogy to the celebrated Vollhardt-Woelfle self-consistent equation theory for Anderson localization\cite{44}. In fact, it has been found recently\cite{45} that upon generalizing this self-consistent equation to topological quantum metals, the diffusion constant turns out to be time dependent and exhibits anomalous quantum diffusion.

\begin{acknowledgments}
We are grateful to J. Wang and C. S. Tian for useful discussions and for a critical reading of the manuscript. This work is supported by the NSFC (Grants No. 11335006).
\end{acknowledgments}

%\bibliography{bibfile}

\end{document}